\documentclass[11pt]{article}
\usepackage{amssymb}\usepackage{amsmath}
\usepackage[dvips]{graphicx}
\usepackage{psfrag}
\renewcommand{\phi}{\varphi}

\newcommand{\beq}{\begin{equation}}
\newcommand{\eeq}{\end{equation}}

\DeclareMathOperator{\RE}{\mathrm{Re}}
\DeclareMathOperator{\IM}{\mathrm{Im}}
\DeclareMathOperator{\e}{\mathrm{e}}
\begin{document}

December 2007  \strut\hfill DESY 07-216 \\
\vspace*{3mm}
\begin{center}
{\large\bf
 On string solutions of Bethe equations \\
 in $\mathcal{N}\,{=}\,4$ supersymmetric Yang--Mills theory
} \\ [3mm]
{\sc Andrei G. Bytsko$^{1,2}$ and 
 \ Igor E. Shenderovich$^{3}$}    \\ [2mm]
{ \small
 ${}^{1}$ Steklov Mathematics Institute,
 Fontanka 27, 191023, St.~Petersburg, Russia \\
 ${}^{2}$ DESY Theory Group, Notkestrasse 85, D--22603 Hamburg, Germany\\
 ${}^{3}$ Physics Department, St.Petersburg State University,
 St.Petersburg, 198504, Russia
} \\ [3mm]
{ }

\end{center}
\vspace{1mm}
\begin{abstract}
The Bethe equations, arising in description of the spectrum 
of the dilatation operator for the $su(2)$ sector of the 
$\mathcal{N}\,{=}\,4$ supersymmetric Yang--Mills theory, are 
considered in the anti--ferromagnetic regime. 
These equations are deformation of those for the Heisenberg
XXX magnet. It is proven that in the thermodynamic limit roots
of the deformed equations group into strings. It is proven that the
corresponding Yang's action is convex, which implies uniqueness of
solution for centers of the strings. The state formed of strings 
of length~$(2n{+}1)$ is considered and the density of their 
distribution is found. It is shown that the energy of such a state
decreases as $n$ grows. It is observed that non--analyticity of the
left hand side of the Bethe equations leads to an additional
contribution to the density and energy of strings of even length.
Whence it is concluded that the structure of the anti--ferromagnetic 
vacuum is determined by the behaviour of exponential corrections to
string solutions in the thermodynamic limit and possibly involves
strings of length~$2$.
\end{abstract}

\section{Introduction}

Integrable models, in particular, spin chains, appear in several 
problems of high energy physics as effective models of interaction. 
For example, the Hamiltonian of the XXX spin chain with the 
non--compact representation of spin $s{=}-1$ arises in description
of scattering of hadrons at high energies~\cite{lipatov_1993,faddeev1},
and also in description of mixing of composite operators under
renormalization in QCD~\cite{derk}.

Mixing of composite operators under renormalization in the 
super--symmetric Yang--Mills theory also gives rise to an 
XXX--chain but with a compact representation. 
In this theory, one considers locally invariant operators 
of the form
\begin{equation}\label{ZW}
 \mathcal{O} = tr (Z^{J_1} W^{J_2} + \mbox{permutations}),
\end{equation}
where $Z$ and $W$ are two complex scalar fields from the 
supermultiplet. The conformal dimensions $\Delta$ of this operators 
comprise the spectrum of the dilatation operator~$D$. 
It is convenient to describe mixing of operators (\ref{ZW}) 
under renormalization with the help of an analogy with 
the quantum spin chain of length $L= J_1 + J_2$, where
each occurrence of $Z$ is represented by a spin up, and 
each occurrence of $W$ is is represented by a spin down. 
For example, the state $ZZZWWZWZ$ corresponds to
the following spin chain 
$\uparrow\uparrow\uparrow\downarrow\downarrow\uparrow %
\downarrow\uparrow\!\!|\rangle$.\ 
An important observation made in \cite{minzar} was that, in the 
$su(2)$--sector of the theory in the one--loop approximation (i.e., 
in the first order in $\lambda = g_{\rm\scriptscriptstyle YM}^2 N$, 
where $g_{\rm\scriptscriptstyle YM}$ is the Yang--Mills coupling 
constant, and $N$ is the number of colours) the dilatation operator $D$ 
can be expressed via the XXX Hamiltonian of spin $s{=}\frac 12$ 
for the described above chain:
\begin{equation}\label{DX}
 D = \text{const} - \lambda \, H_{\scriptscriptstyle \rm XXX}
 + O(\lambda^2) \,.
\end{equation}
Therefore, determining the spectrum of $D$ in this approximation
is reduced to investigating the Bethe equations for the 
Heisenberg magnet (see \cite{fadtak,faddeev2,bik}),
\begin{equation}\label{bethe_eq0}
 \left( \frac{u_j+i/2}{u_j-i/2} \right)^L =
 \prod_{k \neq j} \frac{u_j-u_k+i}{u_j-u_k-i} \,,
\end{equation}
where $u_i$ are the rapidities of elementary excitations.

The high--loop corrections to (\ref{DX}) were found in articles that 
followed \cite{minzar}, and there it was shown that the corresponding 
expressions are also integrable Hamiltonians for the spin chain 
(those that include interaction between several nearest sites). 
In these approximations, the spectrum of $D$ is determined not 
by equations (\ref{bethe_eq0}) but by their ``deformations'' 
which explicitely contain the parameter $\lambda$ in the 
left--hand side. Assuming that integrability of $D$ takes place 
in all orders, Beisert, Dippel, and Staudacher \cite{beisert} 
argued that the exact Bethe equations determining the
spectrum of $D$ look like following
\begin{equation}\label{bethe_eq}
 \left( \frac{x(u_j+i/2)}{x(u_j-i/2)} \right)^L =
 \prod_{k \neq j} \frac{u_j-u_k+i}{u_j-u_k-i} \,,
\end{equation}
where
\begin{equation}\label{xu}
 x(u) = \frac12 \left(u + \sqrt{u^2 - \chi^2}\right) \,, \qquad
 \chi^2 \equiv \frac{\lambda}{4 \pi^2} \,.
\end{equation}

For the spin chain of length $L$, the spectrum of the Hamiltonian 
is bounded by the energies of the ferro\-magnetic and the 
anti\-ferromagnetic vacua. Therefore, the spectrum of the operator $D$, 
i.e. the dimensions of operators (\ref{ZW}), belongs to the interval  
\begin{equation}
 L \leqslant \Delta \leqslant \Delta_{\rm max} \,.
\end{equation}

The lower bound here follows obviously from the structure 
of the anti\-ferro\-magnetic vacuum (all spins up), whereas
determining the upper bound requires a non--trivial evaluation 
of the energy of the anti\-ferro\-magnetic vacuum. 
The value of $\Delta_{\rm max}$ in the thermodynamic limit 
was found in \cite{zarembo} and \cite{staudacher} by means of 
the standard technique of passing to the limit $L \to \infty$ 
in the Bethe equations. In this procedure it was assumed that,
like for equations (\ref{bethe_eq0}), the anti\-ferro\-magnetic 
vacuum is formed of strings of length~1, i.e., real roots of 
equations~(\ref{bethe_eq}). However, the example of the XXX spin 
chain of higher spin [11,12] shows that the anti\-ferro\-magnetic 
vacuum for Bethe equations whose left--hand side differs {}from 
(\ref{bethe_eq0}) can have other structure, for instance,
it can be filled by strings of greater length.

The aim of the present work is to prove existence of solutions 
corresponding to strings of length greater than~1 for the 
equations (\ref{bethe_eq}) in the thermodynamic limit and to check 
validity of the assumption about the structure of 
the anti\-ferro\-magnetic vacuum.

\section{Existence of string solutions}

To make from $x(u)$ an analytic single--valued function, we fix in 
$\mathbb{C}$ the cut $[-\chi,\chi]$ and define $x(u)$ on 
$\mathbb{C}\slash[-\chi, \chi]$ as follows: 
\begin{equation}\label{7}
 x(u) = \frac14 \left( \sqrt{r_1} \, e^{\frac{i}{2} \theta_1} +
 \sqrt{r_2} \, e^{\frac{i}{2} \theta_2} \right)^2, 
\end{equation}
where $r_1,\,r_2 \in \mathbb{R_{+}}$ and $\theta_1,\,\theta_2
\in\, ]{-}\pi,\pi]$ are determined from the relations
$u = \chi + r_1 e^{i \theta_1} = -\chi + r_2 e^{i \theta_2}$.
Notice that the signs of the imaginary parts of $u$ and $x(u)$ 
coincide, that is $x(u)$ maps a point from the upper/lower half--plane 
to the upper/lower half--plane, respectively.

Let us prove that the following relations
\begin{equation}\label{8}
 \left| \frac{x(u+i \delta)}{x(u-i \delta)} \right| \quad 
\begin{cases}
    >1, & \text{for \ $\IM u >0 $} ,\\
    =1, & \text{for \ $\IM u =0 $} ,\\
    <1, & \text{for \ $\IM u <0 $} .
\end{cases}
\end{equation}
hold for every $\delta >0$.

Let us denote $s=\RE(u)$, $t=\IM(u)$, $a=\RE\bigl(x(u)\bigr)$,
$b=\IM\bigl(x(u)\bigr)$. It follows from (7) that $|x(u)|$ is 
a function continuous in $u$  (in particular, even when $u$
crosses the cut), and $|x(\overline{u})|=|x(u)|$. Therefore,
with $s$ being fixed, the function $|x(s+it)|$ is continuous and 
symmetric in~$t$. We will prove that this function is 
convex, of which (8) will  then be an obvious consequence.

It follows {} from (5) that the function inverse to $x(u)$ 
is given by $u(x)=x+\frac{\chi^2}{4x}$. Whence it is easy
to derive the following relations
\begin{equation}\label{9}
 s= \Bigl(1+\frac{\chi^2}{a^2+b^2}\Bigr) \, a , \quad
 t= \Bigl(1-\frac{\chi^2}{a^2+b^2}\Bigr) \, b . 
\end{equation}
Let $\partial_t$ denote the partial derivative w.r.t~$t$
(i.e. $\partial_t s =0$). Applying it to (9) and solving the
system of equations for $\partial_t a$ and $\partial_t b$, 
we find 
\begin{equation}\label{10}
 \partial_t a = \frac{2 \chi^2 ab}{D}, 
 \quad \partial_t b = \frac{1}{D} \,
 \left( (a^2+b^2)^2 + \chi^2 (b^2-a^2) \right), 
\end{equation}
where $D=((\chi-a)^2+b^2)((\chi+a)^2 + b^2)$. Whence we obtain
\begin{equation}\label{11}
 \partial_t |x(u)|^2 = 
	2 \, \frac{a^2+b^2}{D} \, (a^2+b^2+ \chi^2) \, b.
\end{equation}
Due to the remark made after equation (7), $t$ and $b$ are of
the same sign. Therefore, expression (11) is positive/negative 
in the upper/lower half--plane, respectively. 
Hence $|x(u)|$ is convex in $\IM(u)$, which completes the
proof of~(8). 

Relations (8) allow us to adapt the analysis of complex roots 
of equations (3) (see~[5,7]) to the case of equations~(4). 
Namely, it follows from (8) that, in the $L\to\infty$ limit,
the absolute value of the l.h.s. of (4) tends to $\infty$ 
when $\IM(u_j)>0$ and to $0$ when $\IM(u_j)<0$. This implies
that r.h.s. has, respectively, a pole or a zero, i.e. there
must exist also the root $u_j-i$ in the first case and 
the root $u_j+i$ in the second case. Thus, like 
in case of the {\rm XXX} magnet, roots of equations (4) group 
in the thermodynamic limit into ``strings'' which are
complexes of the form $u_{j,m}=u_j+im$, 
where $u_j\in\mathbb{R}$ and $2m\in\mathbb{Z}$.  

\section{Strings of odd length}

\subsection{Equation for the centers of strings}

Now we will investigate what state has the maximal energy 
in the case when the vacuum is filled with strings of length 
$2n+1$, where $n$ is integer. The number of these strings  
$\nu_n$ is fixed by the condition $(2n+1) \nu_n = L/2$. 
Following [5--7], we multiply Bethe equations (4) along 
a string of length $2n+1$. Since the right--hand side of these 
equations is the same as in the ``undeformed'' Bethe equations, 
strings will have the same form, i.e. $u_j = u_j^n + i m$, 
$m \in \mathbb{Z}$. Further, considering the thermodynamic limit 
($L\to\infty$), we obtain that the centers of strings are arranged
along the real axis with some density which satisfies certain
integral equation. Having found this density, one can compute 
the energy of the ground state (see~[5,6]).

Thus, we obtain the following equation for the centers of strings:
\begin{equation}\label{12}
\frac i2 L \ln \frac{x\left(u_j^n
+(2n+1)\frac{i}{2}\right)}{x\left(u_j^n
-(2n+1)\frac{i}{2}\right)} = \pi Q_j^n +
\sum_{k=1}^{\nu_n} \Phi_{n,n} (u_j^n - u_k^n), 
\end{equation}
where
\begin{equation}\label{13}
 \Phi_{n,n}(u) =  \arctan\frac{u}{2n+1} + 2
\sum_{m=0}^{2n-1} \arctan\frac{u}{m+1}. 
\end{equation}

\subsection{Yang's action}

To prove the uniqueness of a solution to equations (12) for
a given set of integer numbers $Q_j^n$, we will use the Yang's action. 
As in the case of the {\rm XXX} magnet \cite{bik}, there exists
a functional $S$ (called Yang's action) such that equations (12) 
are the conditions of its extremum, $\partial_{u_\alpha} S=0$. 
Let us consider the quadratic form for the matrix of second 
derivatives of~$S$: 
\begin{equation}
 \sum_{\alpha,\beta} v_\alpha \,
 \frac{\partial^2 S}{\partial_{u_\alpha} 
 \partial_{u_\beta}} \, v_\beta =
\frac i2 L\sum_\alpha \partial_{u_\alpha} 
\ln \frac{x(u_\alpha + \frac{i}{2}(2n+1))}
 {x(u_\alpha - \frac{i}{2}(2n+1))} \, v^2_\alpha 
\label{14}
  +\! \sum_{\alpha > \beta} \frac{1}{(u_\alpha -
u_\beta)^2+1}(v_\alpha - v_\beta)^2,
\end{equation}
where $v_\alpha \in \mathbb{R}$.
It is obvious that the second term is always positive. 
Let us prove the positivity of the first one:
\begin{equation}\label{15}
\frac{i}{2} \partial_s \log \frac{x(s+it)}{x(s-it)} =
\partial_s \arctan \frac{a}{b} =
\frac{b\partial_s a-a\partial_s b} {(a^2+b^2)}  = 
\frac{a \partial_t a + b \partial_t b}{(a^2+b^2)} =
 \frac{\partial_t|x(s+it)|^2}{2(a^2+b^2)}.
\end{equation}
Here we used the same notation as in Section~2 and applied the 
Cauchy equations for derivatives of an analytic function. 
Relation (11) shows that (15) is positive for $b>0$. 
Therefore the quadratic form (14) is positive definite and, 
consequently, the action $S$ has a unique minimum.

\subsection{Thermodynamic limit}

Let us go to the thermodynamic limit now. Taking $L\to \infty$ and 
differentiating (12) with respect to $u$, we obtain for the 
left--hand side:
\begin{equation}\label{16}
l.h.s. =\frac i2 \bigg(\frac{1}{\sqrt{\big(\frac{(2n+1)}{2}i+u\big)^2
-\chi^2}}-\frac1{\sqrt{\big(\frac{(2n+1)}2i-u\big)^2-\chi^2}}\bigg).
\end{equation}
Now we introduce the root density $\rho(u)$:
\begin{equation}\label{17}
 \rho(u) = \frac{1}{\left(\dfrac{du}{dq}\right)_{q=q(u)}},
\end{equation}
where $q(u)=Q_j/L$. $\rho(u)$ plays the role of density of 
numbers $q(u)$ on the interval~$du$. Having introduced this
density, we can rewrite the l.h.s. of (12) as follows
\begin{equation}\label{18}
r.h.s. = \pi \rho(u) \int\limits_{-\infty}^{\infty} d
\mu\, \rho(\mu)\bigg[ \frac{2n+1}{(2n+1)^2 + (u - \mu)^2 } 
 + 2 \sum_{m=0}^{2n-1} \frac{m+1}{(m+1)^2 + (u - \mu)^2 }\bigg]. 
\end{equation}
This integral equation can be solved by means of the Fourier 
transform. To this end we compute first
\begin{align}
\nonumber
 \int_{-\infty}^{\infty} & du\, 
 \frac{\e^{iku}}{\sqrt{(u+il)^2- \chi^2}} 
 = \theta (-kl) \e^{-|kl|} \oint \frac{dq}{q}
\exp\left(ik \left(q+\frac{\chi^2}{4q}\right)\right) \\ 
\label{19}
{}&=\ \text{\rm sign}\, (-l) \theta (-kl) \e^{-|kl|}
\int_{0}^{2\pi} i d \phi \e^{ik \chi \cos \phi} 
= 2\pi i\, \text{\rm sign}\,(-l) \theta (-kl) \e^{-|kl|} J_0(\chi k), 
\end{align}
where $J_0(k)$ is the Bessel function of the first kind. 
Thus, after the Fourier transform, the left--hand side 
of our integral equation acquires the following form
\begin{equation}\label{20}
 F[l.h.s] = \pi \e^{-\frac{2n+1}{2}|k|} J_0(\chi k) \,.
\end{equation}
On the r.h.s. the Fourier integral is divided into the
following terms
\begin{equation}\label{21}
 \int\limits_{-\infty}^{\infty} d\mu\, \rho(\mu)
\int\limits_{-\infty}^{\infty} \e^{i k u} \frac{A}{A^2 +
(u-\mu)^2} du 
 = \pi \int\limits_{-\infty}^{\infty} d\mu\,
\rho(\mu) \e^{i \mu k} \e^{-|k| A} = \pi \e^{-|k|A}
\widetilde{\rho}(k), 
\end{equation}
where $\widetilde{\rho}(k)$ stands for the Fourier transform 
of the density~$\rho(u)$. These terms can be summed up: 
\begin{equation}\label{22}
\pi \widetilde{\rho}(k) \left(2\sum_{m=0}^{2n-1} \e^{-|k|
(m+1)} + \e^{-|k| (2n+1)} \right)\\ = \pi \widetilde{\rho}(k)
\left( \frac{2-\e^{-2n|k|} - \e^{-(2n+1)|k|}}{\e^{|k|}-1}
\right). 
\end{equation}

As a result, we obtain the following expression for the 
Fourier transformed density $\widetilde{\rho}(k)$:
\begin{equation}\label{23}
\widetilde{\rho}(k)= 
J_0 (\chi k)
\frac{\e^{|k|}-1}{\e^{-\frac{2n+1}{2}|k|}(1+\e^{|k|})(\e^{(2n+1)|k|}-1)}
= \frac{J_0(\chi k)
\tanh\frac{|k|}{2}}{2 \sinh \left(n+\frac12\right)|k|}. 
\end{equation}
This yields the solution of the integral equation,
\begin{equation}\label{24}
 \rho(u) = \frac{1}{2\pi} \int\limits_{-\infty}^{\infty}
\frac{ J_0(\chi k) 
\tanh\frac{|k|}{2}}{2 \sinh \left(n+\frac12\right)|k|}
\e^{iku} dk. 
\end{equation}
For $n=0$ (i.e. strings of length~1) this expression coincides with 
the expression for the density that was obtained in~[9, 10]. 

\subsection{Energy of strings of odd length}

Now we will compute the energy (or, equivalently, the
maximal dimension of the dilatation operator $\Delta_{\max}$) for
the state filled with strings of length $2n+1$, where $n$ is integer. 
The dispersion in the considered theory differs from that of 
the XXX magnet, the corresponding energy density is given by~[9]
\begin{equation}\label{25}
 \frac{\Delta_{\max}}{L} = 1 + \frac{i \lambda}{8\pi^2}
\int\limits_{-\infty}^{\infty} du\, \rho(u) \left(
\frac{1}{x\left(u + \frac{2n+1}{2}i \right)} -
\frac{1}{x\left(u - \frac{2n+1}{2}i\right)} \right). 
\end{equation}

Substituting here the expression (24) for the density, we obtain:
\begin{equation}\label{26}
 \frac{\Delta_{\max}}{L} = 
1 + \frac{\sqrt{\lambda}}{\pi}
\int\limits_0^{\infty} \frac{dk}{k} \frac{J_0(\chi k)
J_1(\chi k)
\tanh\frac{k}{2} }{\e^{(2n+1)k}-1}.
\end{equation}
Here we used the following relation:
\begin{equation}\label{27}
 \frac{\chi^2}{2} \left( \frac{i}{\sqrt{(u-il)^2-\chi^2}}
- \frac{i}{\sqrt{(u+il)^2-\chi^2}} \right) = \frac12 \chi
\partial_{\chi} \left[ \frac{\chi^2}{2i} \left(
\frac{1}{x(u+il)} - \frac{1}{x(u-il)} \right) \right]. 
\end{equation}
Below we provide a plot which shows how the second term 
in (26) depends on $n$ for a fixed value of~$\lambda$.
\begin{center}
\includegraphics[scale=0.8]{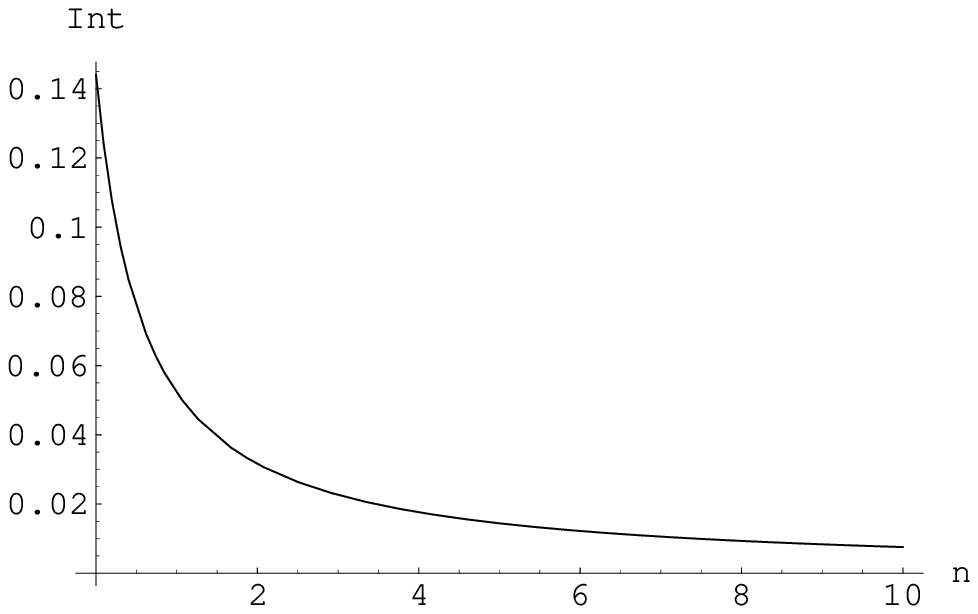}
\end{center}
It is apparent that the maximum of the integral (and thus
the maximum of $\Delta_{\max}$) is attained at~$n=0$. 
In the Appendix we give a strict proof that $\Delta_{\max}$
decreases monotonously as $n$ grows.

Thus, we conclude that the state with maximal energy in the 
sector of strings of odd length indeed corresponds 
to strings of length~1, as it was assumed in~[9, 10]. 
However, in the next section we will show that the real
anti--ferromagnetic vacuum might have a more complicated
structure and possibly corresponds to strings of length~2.

\section{Strings of even length}

Unlike the absolute value $|x(u)|$, the phase of $x(u)$
is not a continuous function. Its value changes by a
finite amount when $u$ crosses the cut. Using (7) it is 
easy to derive that for $u\in [-\chi,\chi]$ we have the
following relation:
\begin{equation}\label{28}
 \lim_{\epsilon  \to 0}
 \ln \frac{x(u-i\epsilon)}{x(u+i\epsilon)} =
 2i \, \nu_\epsilon \, \arctan \frac{\sqrt{\chi^2-u^2}}{u},
\end{equation}
where $\nu_\epsilon =-1$ if $\epsilon$ tends to the zero {}from
the right and $\nu_\epsilon=1$ if $\epsilon$ tends to the zero
{}from the left.

For a chain of a large but finite length $L$, the roots of Bethe
equations group into strings only up to exponential deviations:
$u_{j,m}=u_j +i(m + \epsilon_m)$, where $m$ is integer or
half--integer and $\epsilon_m = O(e^{-\omega_m L})$, $\omega_m>0$.
The total energy of a string is given by
$\mathcal E_j=\frac{i}{2}\sum\limits_{m}
 \Bigl( \frac{1}{x(u_{j,m} + \frac{i}{2})}-
 \frac{1}{x(u_{j,m} - \frac{i}{2})}\Bigr)$.
Computing this expression for a string of even length and taking
relation (28) into account, we observe that the roots
\begin{equation}\label{29}
 u_j +i\Big(\frac{1}{2} {+} \epsilon\Big) , \quad
 u_j -i\Big(\frac{1}{2} {+} \epsilon\Big)
\end{equation}
give additional contributions,
\begin{equation}\label{30}
 {\mathcal E}_j =
 \lim_{\epsilon \to 0} \frac{i}{2} \bigg(
 \frac{1}{x(u_j {-} i \epsilon)} - \frac{1}{x(u_j {+} i \epsilon)}+\\
 +  \frac{1}{x (u_j {+} \frac{2n+1}{2}i ) } -
 \frac{1}{x(u_j {-} \frac{2n+1}{2}i )} \bigg).
\end{equation}
It is important to remark here that, for the regime corresponding
to $\nu_\epsilon=-1$, the first two terms in (30) give a
negative contribution to~${\mathcal E}_j$. Furthermore,
it can be shown that in this regime ${\mathcal E}_j$ is positive
not everywhere on the real axis. As a consequence, the 
anti--ferromagnet vacuum in this case has a more complicated
structure --- it has to be filled with strings only in
the intervals where ${\mathcal E}_j>0$. In the present work
we will consider only the regime corresponding to 
$\nu_\epsilon=1$. Let us remark that in this case the 
proof of convexity of the Yang's action given in Section~3.2
remains valid.

Taking product of the left hand sides of the Bethe equations (4)
along a string of even length and taking into account the 
additional contributions due to the roots (29), we find that
the l.h.s. of the equation for the centers of strings 
looks like following
\begin{equation}\label{31}
 \lim_{\epsilon  \nearrow 0}
 \left(  \frac{x(u_j-i\epsilon)}{x(u_j+i\epsilon)} \right)^L \,
 \left( \frac{x(u_j+\frac{i}{2}(2n+1))}%
 {x(u_j-\frac{i}{2}(2n+1))} \right)^L .
\end{equation}
Taking logarithm of this expression, we obtain equations~(12)
but with an additional term on the left hand side.
Differentiating (28) w.r.t. $u$, then making the 
Fourier transformation,
\begin{equation}\label{32}
 F\Bigl[ \partial_u \lim_{\epsilon \to 0}
 \ln \frac{x(u-i\epsilon)}{x(u+i\epsilon)} \Bigr]=
 - 2i \, \nu_\epsilon\,
 \int_{-\chi}^{\chi} \, \frac{e^{iku} \,du}{\sqrt{\chi^2-u^2}}
 = - 2\pi i \, \nu_\epsilon\, J_0(\chi k) ,
\end{equation}
and comparing with formula (19), we see that, for strings of
even length in the regime $\nu_\epsilon=1$, equation (20) 
acquires the following form
\begin{equation}\label{33}
 F[l.h.s] = \pi \bigl( e^{-\frac{2n+1}{2}|k|}
 + 1 \bigr)  J_0 (\chi k ) .
\end{equation}
Repeating now the computation of the Fourier image of the 
density as was done in Section~3.3, we find 
\begin{equation}\label{34}
 \rho_\epsilon(u) =
 \frac{1}{2\pi} \int\limits_{-\infty}^{\infty}
 \frac{ J_0 (\chi k)
 \bigl(1 + e^{\frac{2n+1}{2}|k|} \bigr)
 \tanh\frac{|k|}{2}}%
 {2 \sinh \left(n+\frac12\right)|k|} \, e^{iku} dk.
\end{equation}

Formula (33) shows that the additional contributions {}from the 
roots (29) yield the same effect as if the vacuum were filled 
with mixture of strings of length $(2n{+}1)$ and strings of length
$0$ (which formally corresponds to $n=-\frac{1}{2}$). In this
context we remark that expressions of the type (31) appear
in the left hand sides of Bethe equations for chains with
alternating spins (see, e.g.~[13, 14]).

Formula (30) implies that, in order to compute the total energy
of the chain, we have to replace (25) with
\begin{equation}\label{35}
 \frac{\Delta}{L} = 1 + \frac{i \chi^2}{2}
 \lim_{\epsilon \nearrow 0}
 \int\limits_{-\infty}^{\infty} du\, \rho_\epsilon(u)
 \left( \frac{1}{x(u - i \epsilon)} -
 \frac{1}{x(u + i \epsilon)}  \right. \\
 + \left. \frac{1}{x\left(u + \frac{2n+1}{2}i \right)} -
 \frac{1}{x\left(u - \frac{2n+1}{2}i\right)}  
  \right) .
\end{equation}
Substituting here (34) and making the same computation in the 
Section~3.4, we obtain
\begin{equation}\label{36}
 \frac{\Delta }{L} =
 1 +  2 \chi \int\limits_0^{\infty}
 \frac{dk}{k} \, J_0 (\chi k) J_1(\chi k) \, \tanh\frac{k}{2} \,
 \coth\frac{(2n{+}1)k}{4}.
\end{equation}
Using the method described in the Appendix, one can show that
for $n \geq \frac{1}{2}$ expression (36) decreases monotonously as 
$n$ grows. The maximal value of (36) at $n=\frac{1}{2}$ is given by
\begin{equation}\label{37}
 \frac{\Delta_{\max}}{L} =
 1 +  2 \chi \int\limits_0^{\infty}
 \frac{dk}{k}  \, J_0 (\chi k) J_1(\chi k) .
\end{equation}
In order to compare this expression with the value of
$\frac{\Delta_{\max}}{L}$ corresponding to strings of length~1,
one can subtract the value of (26) for $n=0$ from~(37). 
The resulting expression has the form of the integral in (38)\
with monotonously decreasing in $k$ function~$f(k)$. 
As shown in the Appendix, such an integral is positive.

Thus, filling the vacuum with strings of length~2 in the regime
$\nu_\epsilon=1$ yields larger value for $\frac{\Delta_{\max}}{L}$
than filling it with strings of length~1. However, what
regime is indeed realized for string of even length in the 
thermodynamic limit, remains at present an open question.
Its resolution requires quite subtle analysis of the 
exponential corrections $\epsilon$ in (29) for~$L \to \infty$.

\section*{Appendix}

Let us denote $\alpha=(2n+1)$, $\chi=\sqrt{\lambda}/(2\pi)$ 
and make in (26) a substitution $k'=\chi k$. Then
\begin{equation}\label{38}
\partial_\alpha \Bigl( \frac{\Delta_{\max}}{L} \Bigr) =
  - 2 \,\int\limits_0^{\infty} dk \,
 J_0 (k) \, J_1(k) \, f(k) ,
\end{equation}
where
\begin{equation}\label{39}
  f(k) = \biggl( \frac{1}{e^{\frac{k}{\chi}} +1} \biggr)\,
 \biggl( \frac{e^{\frac{\alpha k}{\chi}} }%
 {e^{\frac{\alpha k}{\chi}} -1} \biggr)\,
 \biggl( \frac{e^{\frac{k}{\chi}} -1}%
 {e^{\frac{\alpha k}{\chi}} -1} \biggr).
\end{equation}
In this form, it is evident that $f(k)$ decreases monotonously
as~$k$ grows for all $\alpha >1$ and $\chi >0$. Using this
we will show that the integral in (38) is positive.

Let $0< t_1 < t_3 <t_5 \ldots$ be the ordered set of roots
of $J_0(t)$, and $0=t_0 < t_2 < t_4  \ldots$ be the ordered set 
of roots of $J_1(t)$. It follows from the relation
\begin{equation}\label{40}
  \partial_t J_0(t) = -J_1(t)
\end{equation}
that $t_{2n} < t_{2n+1} < t_{2n+2}$. Taking into account that
$J_0(t) J_1(t)$ is positive on $]t_{2n},t_{2n+1}[$ and negative on
$]t_{2n+1},t_{2n+2}[$, and the function $f(k)$ is positive and
monotonously decreasing for all $k>0$ , we obtain the following
estimate:
\begin{align}
\label{41}
 \int\limits_0^{\infty} dk &\ J_0 (k) \, J_1(k) \, f(k) =
 \sum_{n=0}^\infty \int_{t_{2n}}^{t_{2n+2}}
 dk \, J_0 (k) \, J_1(k) \, f(k) > \\
\nonumber
{}& > \sum_{n=0}^\infty \Bigl[ f(t_{2n+1})  
 \int_{t_{2n}}^{t_{2n+2}} dk \, J_0 (k) \, J_1(k) \Bigr] 
=\frac12 \sum_{n=0}^\infty \Bigl[ f(t_{2n+1})  \, \bigl(
 J^2_0 (t_{2n}) - J^2_0(t_{2n+2}) \bigr) \Bigr] .
\end{align}
In the last equality we used relation~(40). Now, since
roots of $J_1(t)$ are the points of local extrema for
$J_0(t)$, and the values of $|J_0(t)|$ at these points
form a decreasing sequence, we conclude that the sum on the
r.h.s. of (41) and hence the initial integral are positive.
Thus, $\partial_\alpha \Bigl( \frac{\Delta_{\max}}{L} \Bigr) <0$,
which implies that $\frac{\Delta_{\max}}{L}$ decreases monotonously
as~$n$ grows.

\par\vspace*{4mm}\noindent
{\bf Acknowledgement.}
 A.B. was supported in part by
 the Humboldt Foundation, by the Russian Science 
 Support Foundation and by the Russian Foundation for 
 Fundamental Research under grant 05--01--00922.
 I.S. was supported in part by the Dynasty Foundation.

 AB is grateful to V.~Schomerus and J.~Teschner
 for useful discussions and for warm hospitality
 at DESY (Hamburg), where a part of this work was done.

\def\baselinestretch{1}

\end{document}